\documentclass[aps,prl,twocolumn,showpacs,superscriptaddress,superscriptaddress]{revtex4-1}  
\usepackage{graphicx,color}  
\usepackage{dcolumn}   
\usepackage{bm}        
\usepackage{amssymb}   

\usepackage[section]{placeins}
\usepackage{float}
\begin{document}

\title{Criticality in the approach to failure in  amorphous solids}

\author{Jie Lin}
\affiliation{Center for Soft Matter Research, Department of Physics, New York University, New York, NY 10003}
\author{Thomas Gueudr\'e}
\affiliation{DISAT, Politecnico Corso Duca degli Abruzzi, I-10129 Torino - Italy}
\author{Alberto Rosso}
\affiliation{Laboratoire de Physique Th{\'e}orique et Mod{\`e}les  Statistiques (UMR CNRS 8626), Universit\'e de Paris-Sud, Orsay Cedex, France }
\author{Matthieu Wyart}
\affiliation{Institute of Theoretical Physics, Ecole Polytechnique Federale de Lausanne (EPFL), CH-1015 Lausanne, Switzerland}

\date{\today}

\begin{abstract}
Failure of amorphous solids is fundamental to various phenomena, including landslides and earthquakes. Recent experiments
indicate that highly plastic regions form elongated structures that are especially apparent near the maximal shear stress $\Sigma_{\max}$ 
where failure occurs. This observation suggested that $\Sigma_{\max}$  acts as a critical point where the length scale of those 
structures  diverges, possibly  causing  macroscopic transient shear bands. Here we argue instead that the entire solid phase ($\Sigma<\Sigma_{\max}$) is critical, that plasticity always involves system-spanning events, and that their magnitude diverges at $\Sigma_{\max}$ independently of the presence of shear bands. We relate the statistics and fractal properties of these rearrangements to an exponent $\theta$ that captures the stability of the material, which is observed to vary continuously with stress, and we confirm our predictions in elastoplastic models.

\end{abstract}
\pacs{}
\maketitle
Amorphous solids, such as emulsions, sand or molecular glasses are yield stress materials: they behave as solids if the applied shear stress $\Sigma$ is low, but flow as fluids if it is large. Unlike the melting transition, the associated phase transition is dynamical:  the solid phase is an arrested, glassy state whose properties depend on preparation. One such property is failure \cite{Manning07}, which occurs as  $\Sigma$ increases toward a history-dependent  stress $\Sigma_{\max}$ where macroscopic flow starts. For densely prepared materials, the stress overshoots: $\Sigma_{\max}>\Sigma_c$ \cite{Andreotti13, Jagla07}, where $\Sigma_c$ is the minimum stress at which flow can be maintained in stationary conditions. Flow then tends to localize along transient (but sometimes long-lasting) shear bands \cite{Divoux15}. By contrast, for loosely prepared materials  $\Sigma_{\max}=\Sigma_c$ \cite{Andreotti13} and shear-banding may be avoided \cite{Moorcroft13}. 
Despite its importance in human applications and geophysical phenomena, including landslides and earthquakes \cite{Lockner02}, 
the microscopic mechanisms controlling plasticity and failure remain debated. 

In granular materials, recent experiments \cite{Amon2012,Lebouil2014,Lebouil14a} and numerics \cite{Gimbert2013} support that for $\Sigma<\Sigma_{\max}$, plasticity occurs via localized rearrangements, or shear transformations \cite{Argon79}, which tend to organize into elongated structures whose magnitude grows as $\Sigma\rightarrow \Sigma_{\max}$. In \cite{Gimbert2013} it was argued that for  a dense initial state  ($\Sigma_{\max}>\Sigma_c$),  $\Sigma_{\max}$ acts as a critical point where a correlation length $\xi$  diverges and avalanches become system-spanning, which may in turn trigger macroscopic shear bands. This viewpoint complements the  growing consensus that the reverse  transition, occurring when flows stop as the stress is decreased toward $\Sigma_c$, is accompanied by a diverging length scale \cite{Pouliquen04,Lin2014,Olsson07,Martens11,During14,Salerno12,Lemaitre09}. 
Such a ``symmetric" scenario where $\xi$ diverges from both sides of the transition applies to the depinning transition \cite{Fisher98} of an elastic manifold  pushed through a disordered medium. Nevertheless, an alternative scenario has been argued for in glassy systems with slowly-decaying interactions, predicting system-spanning avalanches ($\xi=\infty$) in the entire glass  phase \cite{Muller14}. Applied to amorphous solids, this view suggests criticality for all  stresses $\Sigma<\Sigma_{\max}$ where plasticity occurs. This approach however lacks empirical support, and its consequences on failure near $\Sigma_{\max}$ have not been investigated. 


In this Letter we show that as the stress is adiabatically increased in the solid phase, leading to a  plastic strain $\epsilon(\Sigma)$, the mean avalanche size $\langle S\rangle$ follows $\langle S\rangle  \sim N^{\frac{\theta}{\theta+1}}/(\partial \Sigma/\partial \epsilon)$, where $N$ is the system size and $\theta$ is an exponent that characterizes the stability of the structure \cite{Lin14a}. This result confirms that avalanches are system-spanning ($\xi=\infty$) for all $\Sigma<\Sigma_{\max}$, and further implies an additional singularity as failure is approached,  since $\partial \Sigma/\partial \epsilon\rightarrow 0$ when $\Sigma\rightarrow \Sigma_{\max}$. We suggest that  data analysis used in the literature can mistakenly interpret this singularity as a diverging length scale. We also derive a scaling relation between $\theta$ and exponents characterizing the statistics of avalanches. We test these predictions using elasto-plastic models \cite{Picard2005, Baret02},  and show that they hold independently of the system preparation and of the presence of shear bands near $\Sigma_{\max}$, thus implying that macroscopic flow localization and singularities in avalanche size are unrelated. 

{\it Elastoplastic viewpoint:} following \cite{Hebraud98,Picard2005, Baret02} we model  amorphous solids as consisting of $N$ blocks, each characterized by a scalar local stress $\sigma_{i}$ and a local failure threshold  $\sigma_{i}^{th}$. The overall shear stress  is $\Sigma=\sum_i \sigma_i/N$. Stability of  $i$ is achieved if $|\sigma_i|< \sigma_{i}^{th}$. Otherwise, the block is unstable: a plastic strain of magnitude $\Delta \epsilon_i$ occurs  on some time scale $\tau_c$, leading to an overall increment of plastic strain $\Delta \epsilon=\Delta \epsilon_i/N$.  Such a  plastic event also reduces stress locally by some amount $\Delta \sigma_i=\mu \Delta \epsilon_i$ where $\mu$ is the elastic modulus, and affects stress in other locations via a long-range Eshelby field $\delta\sigma_j= G(\vec {r}_{ij}) \Delta \sigma_i$ \cite{Picard04}, which can in turn trigger new instabilities. For our numerics below, we choose the specific model described in \cite{Lin2014} in two dimensions. Blocks then form a bi-periodic square lattice, and the elastic propagator follows approximatively ${\cal G}({\vec r}_{ij})\propto \cos (4\phi)/r^2$ where $\phi$ is the angle between the shear direction and ${\vec r}_{ij}$. We choose $\sigma^{th}=\tau_c=\mu=1$, and  $\Delta \sigma_i=-\sigma_i+\delta $, where $\delta$ is a random number, uniformly distributed in $[-0.1, 0.1]$. For these choices,  $\Sigma_c \simeq 0.53$, and stability is easily expressed in terms of the variable $x_i\equiv \sigma_{i}^{th}-\sigma_i$, and corresponds to  $x_i\in[0,2]$. 

\begin{figure}[htb!]
   \includegraphics[width=.48\textwidth]{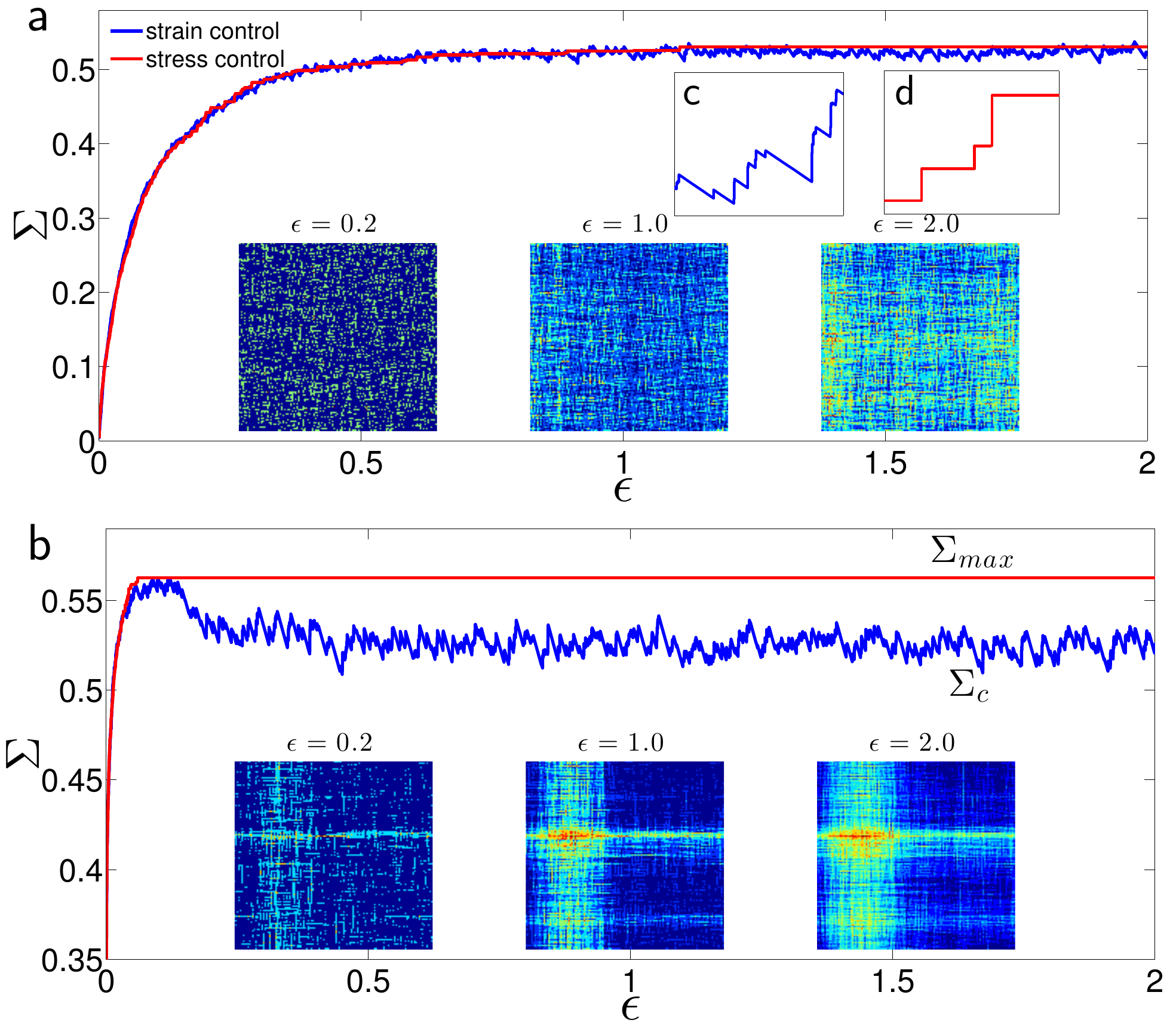} 
    \caption{Stress $\Sigma$ {\it v.s.} plastic strain $\epsilon $ curves for both strain (blue) and stress (red) controlled protocols, for (a) a broad initial distribution $P_0(x)$ and (b) a narrow $P_0(x)$. In (a) $\Sigma_{\max}=\Sigma_c$, whereas in (b) the stress overshoots and $\Sigma_{\max}>\Sigma_c$.  Insets: spatial maps of plastic strain measured at different strain level.  Highly (weakly) plastic regions are indicated in yellow (blue). Macroscopic shear localization occur in (b) but not in (a). (c,d): Zooming in on the  stress {\it v.s.} plastic strain curves, one observes microscopic differences between the two protocols. }\label{f1}
\end{figure}

Such automaton models can be used to study the transient regime  toward failure. In what follows we use two quasi-static protocols. In the stress-control protocol, $\Sigma$ is increased just sufficiently to trigger a single instability. $\Sigma$ is fixed during the resulting avalanche, and is increased again only when this chain of events has stopped. The strain-control protocol is identical, except that $\Sigma$ decreases during avalanches, proportionally to the plastic strain. Stress {\it v.s.} plastic strain curves for these two protocols are shown in Fig.\ref{f1} (from which the stress {v.s.} total strain $\gamma$ curves  are  easily deduced using the relation $\Delta \gamma= \Delta\epsilon +\Delta \Sigma/\mu$). They essentially track each other macroscopically (although they differ microscopically, see insets (c) and (d)) except when $\Sigma$ reaches $\Sigma_{\max}$, if $\Sigma_{\max}>\Sigma_c$. 

The transient qualitatively depends on the initial stability of the system, characterized by the initial distribution of local stability $P_0(x)$. If  $P_0(x)$ is narrow and depleted near $x=0$ (corresponding to a very stable initial condition), transient shear bands occur; otherwise flow can remain homogeneous \cite{Vandembroucq2011}. In Fig.\ref{f1} we confirm these results using a broad and a narrow distribution $P_0(x)$ (see S.I. for details). We further find that transient shear bands tend to occur if the stress-strain curve overshoots (although we did not investigate this correlation systematically), as is sometimes reported \cite{Shi2007, Moorcroft2011,Jagla07} and argued for in \cite{Moorcroft13,Fielding2014}. In what follows we focus on avalanche-type response, for $\Sigma$ below and approaching $\Sigma_{\max}$.


\begin{figure}[htb!]
   \includegraphics[width=.5\textwidth]{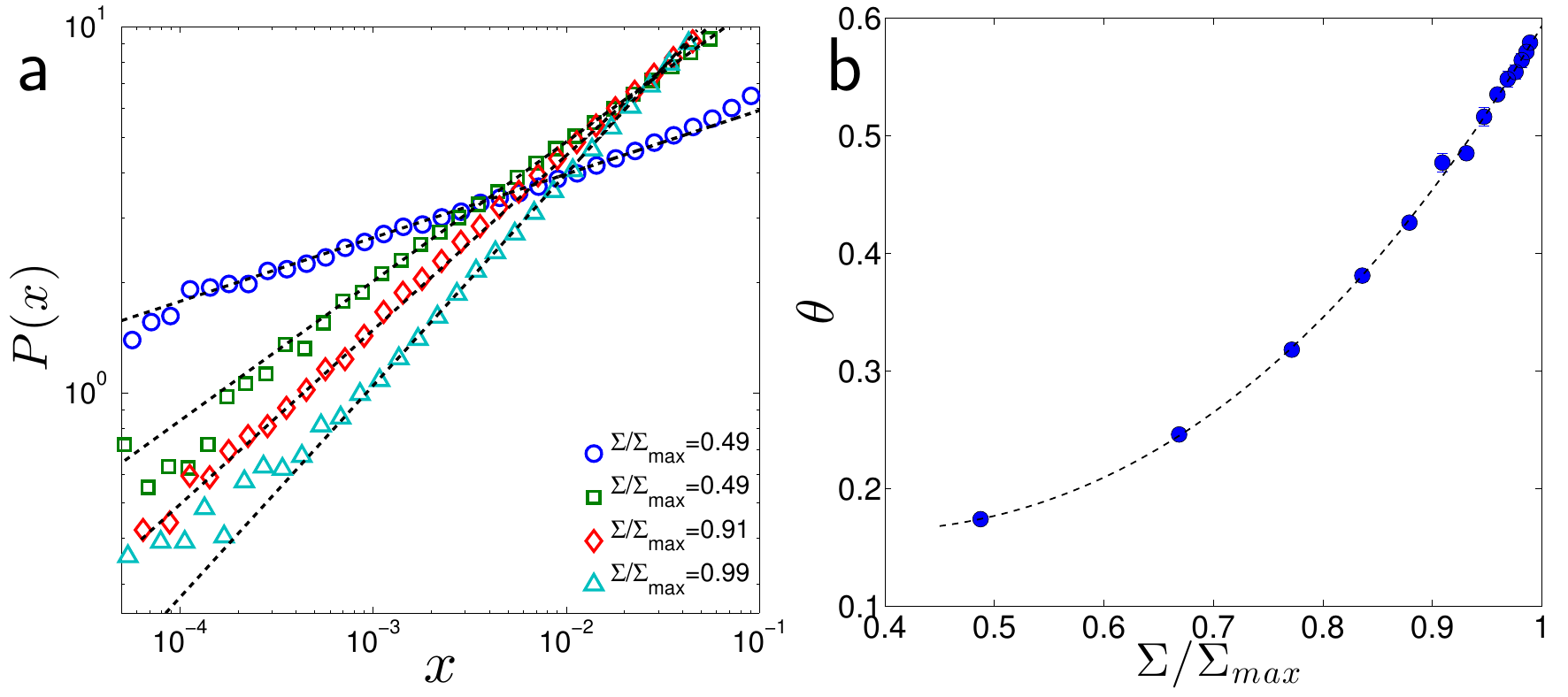} 
    \caption{(a) Distribution of local stability $P(x)$ for $\Sigma/\Sigma_{max}=0.49$ to $0.99$
in the stress-control protocol for $N=1024^2$ in the case $\Sigma_{\max}=\Sigma_c$. The dashed lines are direct fits of the form $P(x)\sim x^\theta$, from which we extract $\theta$.  This quantity is shown in (b). The dashed line is the interpolation using a third order polynomial fit. }\label{f2}
\end{figure}

{\it Distribution of local distance to yield stress:} mean-field  models \cite{Lemaitre07,Hebraud98} reveal that the distribution of local stability $P(x)$ vanishes near $x=0$ in a quasi-static shear at $\Sigma_c$. In \cite{Lin14a} some of us showed that stability indeed requires the presence of a {\it pseudo-gap}, i.e. $P(x)\sim x^\theta$ with $\theta>0$, otherwise, any plastic event would eventually trigger an extensive rearrangement, and this argument also holds in the transient regime.  $\theta$ was measured in elasto-plastic models \cite{Lin14a} and indirectly in  MD simulations \cite{Karmakar10a,Salerno13} both at $\Sigma_c$ and after a quench at $\Sigma=0$, leading to consistent results. In Fig.\ref{f2} we extend these results to the transient regime. We find that $\theta>0$ as predicted in \cite{Lin14a}. However, the value of $\theta$ turns out to be function of the relative stress $\Sigma/\Sigma_{max}$, while it converges to a well-defined value for large system size as shown in S.I. After some initial decay at very small $\Sigma$ (not shown), the value of $\theta$ increases from $\theta=0.174\pm0.004$ at $\Sigma/\Sigma_{max}\approx 0.49$ to the value $\theta=0.6\pm0.004$ at $\Sigma=\Sigma_{max}$. This measure is consistent with the exponent obtained in the stationary regime  \cite{Lin2014}. 


The value of $\theta(\Sigma_c)$  was argued to control rheological properties in the flowing phase as $\Sigma\rightarrow \Sigma_c$ from above\cite{Lin2014} and to imply system spanning avalanches for $\Sigma<\Sigma_{max}$ \cite{Muller14}. We now extend this latter argument to  include the case where $\Sigma\rightarrow\Sigma_{max}$ from below.

Extreme value statistics implies that if $P(x)\sim x^\theta$ and the variables $x_i$ are independent, the least stable block must be at a distance $x_{min}\sim N^{-\frac{1}{\theta+1}}$ of an instability. By definition, $x_{min}$ is the increment of stress that can be added before a new avalanche starts: the length of the  vertical lines in Fig.\ref{f1}(d). 
Hence, following a finite stress increment of $\Delta \Sigma$, a number $M \sim \Delta \Sigma / x_{\min}\sim\Delta \Sigma N^{\frac{1}{\theta+1}}$ of avalanches are triggered. In elastoplastic models the avalanche size $S$ is defined as the number of plastic events, which is approximately related to the total strain of the single avalanche $\delta\epsilon$, by $S\approx N\delta \epsilon$. 
Thus the total strain increase $\Delta\epsilon$ must follow $\Delta\epsilon=M\langle \delta \epsilon\rangle =M\langle S\rangle /N$, where $\langle S\rangle $ is the mean avalanche size at stress $\Sigma$. We thus get:

\begin{equation}
\label{01}
\langle S\rangle=\frac{N \Delta \epsilon}{M}=\frac{N^{\theta/(1+\theta)} \Delta \epsilon}{\Delta \Sigma}\rightarrow  \frac{N^{\theta/(1+\theta)}}{\partial \Sigma/\partial\epsilon}\label{eq1}
\end{equation}
%
where $\partial \Sigma/\partial\epsilon$ is the local slope of the stress-plastic strain curve, and the limit corresponds to $\Delta \Sigma\rightarrow 0$.
This central result indicates that (i) if $\Sigma$ is increased in the solid phase, avalanches are system-spanning ($\xi=\infty$) even for $\Sigma<\Sigma_{max}$, since their size is $N$ dependent. Thus  the system remains critical in the whole range $ \Sigma < \Sigma_{max}$ as long as plastic flow occurs, i.e. $\partial \Sigma/\partial\epsilon<\infty$.  (ii) Avalanches become larger as $\Sigma\rightarrow \Sigma_{\max}$, as observed \cite{Lebouil2014}, since $\partial \Sigma/\partial\epsilon\rightarrow 0$ at $\Sigma_{\max}$. 

Further scaling relations can be derived for the statistical properties of  transient avalanches for $\Sigma<\Sigma_{\max}$. We make the assumption that the distribution of avalanches $P(S)$ is homogeneous, i.e. $P(S)=S^{-\tau}f(S/S_c)$, where the cut-off size scales  as $S_c\sim L^{d_f}$. Here $d_f$ is the fractal dimension of avalanches, $L$ is the linear system size, and $N=L^{d}$, where $d$ is the spatial dimension. From this distribution it is straightforward to compute the mean $\langle S\rangle\sim L^{d_f(2-\tau)}$. Comparing with Eq.(\ref{01}), we get:
\begin{equation}
	\tau=2-\frac{d}{d_f}\frac{\theta}{\theta+1}\label{tau}
\end{equation}
A similar relation holds for stationary flow \cite{Lin2014}, although in the transient regime exponents appear to depend continuously on $\Sigma$. 

Finally, we introduce an exponent $\gamma$ defined as $d\Sigma/d\epsilon\sim(\Sigma_{\max}-\Sigma)^{\gamma}$ for $\Sigma$ close to $\Sigma_{max}$. Eq.(\ref{01}) then implies the scaling relation: 
\begin{equation}
\langle S\rangle \sim (\Sigma_{\max}-\Sigma)^{-\gamma} N^{\frac{\theta}{\theta+1}}\label{gamma}
\end{equation}

%

\begin{figure}[htb!]
   \includegraphics[width=.5\textwidth]{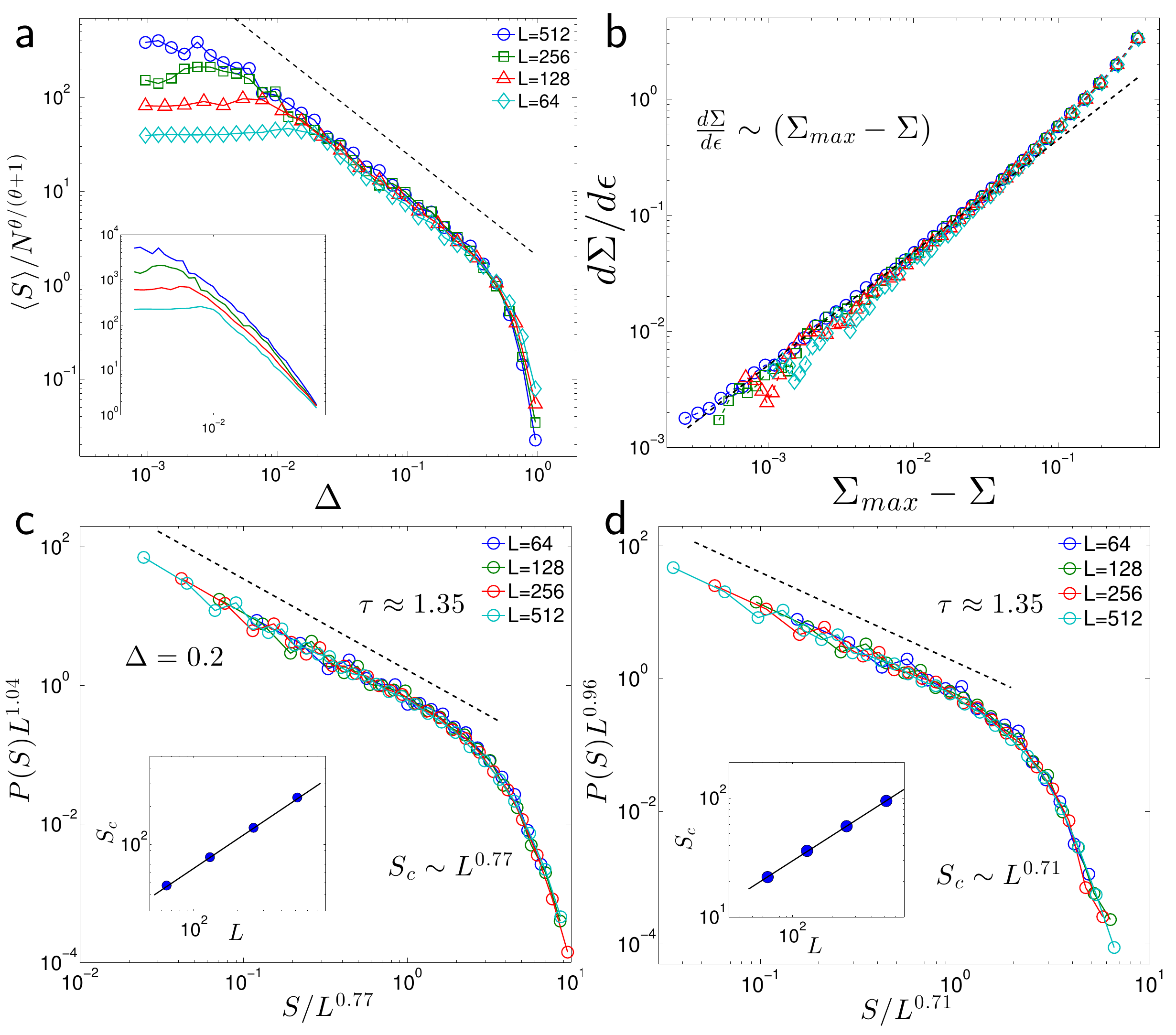} 
    \caption{(a) Collapse of the mean avalanche size  as function of the proximity to failure $\Delta\equiv\frac{\Sigma_{\max}-\Sigma}{\Sigma_{\max}}$,
 using the value of $\theta(\Sigma)$ extracted from Fig.\ref{f2}(b). The slope of the dotted line is $-1.1$.
Here $\Sigma_{max}$ depends on the system size. The inset is the same quantity with no rescaling. 
(b)  Local slope $d\Sigma/d\epsilon$ {\it vs} $\Sigma_{\max}-\Sigma$, supporting $\gamma\approx 1$, corresponding to $d\Sigma/d\epsilon\sim (\Sigma_{max}-\Sigma)$ asymptotically. (c)\&(d) Collapse of the distribution of avalanche size at a specific stress value corresponding to $\Delta=0.2$
for the stress-control case(c)  and strain-control case(d). 
We get $d_f\approx 0.77(0.71)$ in the stress(strain)-control case, and $\tau\approx1.35$ in both cases. All  numerics are for $d=2$.
}\label{f3}
\end{figure}

These predictions are tested in Fig.\ref{f3}. The inset of panel (a) shows that the mean avalanche size, as a function of $\Delta=\frac{\Sigma_{max}-\Sigma}{\Sigma_{max}}$, grows with the system size even far from failure. The entire solid phase is critical, as expected from Eq.(\ref{eq1}). Note that to test this equation, one must consider the fact that $\theta=\theta(\Sigma)$. In this figure we use for $\theta(\Sigma)$ the third order polynomial fit of Fig.\ref{f2}(b). Using these values for $\theta$ a beautiful collapse is observed.

The presence of system sized avalanches far from threshold has to be distinguished from the divergence observed close to the yield stress, $\langle S\rangle\sim (\Sigma_{\max}-\Sigma)^{-\gamma}$ at fixed $N$ as implied by Eq.(\ref{gamma}). Fig.\ref{f3}(a) is consistent with this relation and yields $\gamma\approx 1.1$.  According to its definition, $\gamma$ can also be directly measured from the local slope of stress-strain curves, as is done in Fig.\ref{f3}(b) where $\gamma\approx 1$ is found, consistent with Fig.\ref{f3}(a). $\gamma=1$ means that the stress tends to $\Sigma_{\max}$ exponentially fast. As shown in S.I, this appears to be valid also if the stress overshoots and $\Sigma_{\max}>\Sigma_c$.

In Fig.\ref{f3}(c,d), we measure $d_f$ at $\Delta=(\Sigma_{\max}-\Sigma)/ \Sigma_{\max}=0.2$, where $\theta\approx 0.33$, by collapsing the probability distribution of avalanche sizes, $P(S)\sim S^{-\tau}f(S/S_c)$ with $S_c\sim L^{d_f}$. We find $d_f\approx 0.77$ and $\tau\approx 1.35$ in the stress-control case. Again, these values perfectly agree with Eq.(\ref{tau}). This result holds also for the strain-control protocol where we  find scale-free avalanches with the same $\tau$ and a similar fractal dimension, $d_f\approx 0.71$. 


\begin{figure*}[htb!]
   \includegraphics[width=.8\textwidth]{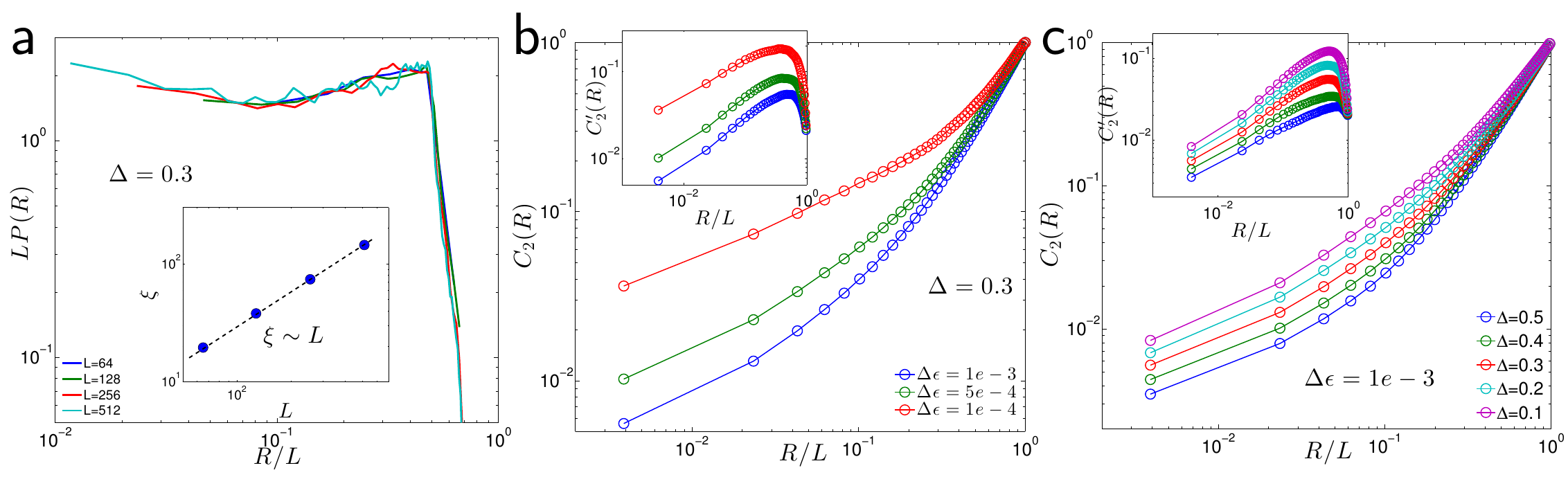} 
    \caption{(a) Distribution of avalanche extension $P(R)$ for the stress-control protocol at  $\Delta=0.3$.  Collapse occur by rescaling distances with $L$, supporting that  $\xi=L$. Inset: direct measurement of $\xi$, defined as $\xi\sim \langle R\rangle$.  (b) $C_2(R)$ at $\Delta=0.3$ for different $\Delta\epsilon$ as indicated in legend,  suggesting a length scale that depends on $\Delta\epsilon$. Inset:  $C^{\prime}_2(R)$ for which no such dependence appears. (c) $C_2(R)$ computed for $\Delta\epsilon=10^{-3}$ and varying $\Delta$ as indicated in legend, suggesting an increasing length scale as $\Delta\rightarrow 0$. Inset: $C^{\prime}_2(R)$ shows no such effect. }\label{f4}
\end{figure*}

{\it Length scale:} To further evidence the presence of a diverging length scale throughout the solid phase, we study the strain map generated by a single avalanche, and consider the $M=(S-1)S/2$ distances $|\vec{R_i}-\vec{R_j}|$ between all the blocks involved in the avalanche. We compute the distribution of these distances, and define $P(R)$ as the average of these distributions among  avalanches occurring at the same stress value in different samples (we choose to weight each avalanche by $M$ in this average). We focus on avalanches occurring at a finite distance from failure, with $\Delta\approx 0.3$. Assuming homogeneity we expect $P(R)=\frac{1}{\xi^\alpha}g \left(\frac{R}{\xi} \right)$. In Fig.\ref{f4}(a), we confirm such a form, specifically:
\begin{equation}
P(R)=\frac{1}{L}g \left(\frac{R}{L} \right)\label{pr}
\end{equation}
We observe a similar scaling form in the strain-control simulation. These results confirm that $\xi\sim L$, as further supported by the observation that $\langle R\rangle\sim L$ shown in inset.


Our results are at odd with the conclusions of \cite{Gimbert2013}, which report an increasing length scale in a stress-control simulation of granular media. We now suggest that their data may in fact be consistent with our views.  In \cite{Gimbert2013}, a length scale is extracted by considering the fluctuations  of the strain field obtained during some strain interval $\Delta\epsilon$, for different stress values $\Sigma$. This is {\it a priori} different from our analysis above which considers avalanches individually. To clarify this point, we perform an analysis closer to theirs, where finite intervals of strain are considered. We define a pair density function \cite{Girard2010} $C_2(R)$ as the probability that two local plastic events among the $M\approx N \Delta\epsilon$ ones in this interval are at a distance smaller than $R$. Fig.\ref{f4}(b) shows $C_2(R)$ for $\Delta=0.3$ and varying $\Delta\epsilon$ as indicated in legend. At first sight, one may think that  
a length scale can be extracted from $C_2(R)$, but that this length is $\Delta\epsilon$-dependent. We find that this dependence, however, can be cured by removing the effect of the mean strain in our definition of $C_2(R)$.  We define $C_2^{\prime}(R)=C_2(R)-C_2(L)R^2/L^2$, which is zero if the plastic events are homogeneous in space. As shown in the inset of Fig.\ref{f4}(b), the characteristic length in $C_2^{\prime}(R)$ does not depend on $\Delta\epsilon$. In Fig.\ref{f4}(c) we show a similar analysis as the proximity to failure $\Delta$ is varied. From $C_2(R)$ it would appear that a length scale grows as $\Delta\rightarrow 0$. However, as shown in inset this is an artefact of this analysis, as $C_2^{\prime}(R)$ shows a constant length scale of order $L$, consistent with our prediction $\xi=\infty$. Experimental measurements of the anisotropic part of the strain field also support that the correlation length is always large and weakly depends on $\Sigma$\cite{Lebouil14a}. Our views could be further tested by performing a similar analysis in similar experimental \cite{Amon2012,Lebouil2014,Lebouil14a} and numerical  \cite{Gimbert2013} data. 

{\it Conclusion:} Ref.\cite{Muller14} argues that glassy systems whose elementary excitations display sufficiently long-range interactions
(including electron glass, mean-field spin glasses or  spheres at random close packing) must display criticality for an entire range of fields or shear stress. This view has not yet been established experimentally. Our work supports that it holds in amorphous solids and granular materials, where it should be testable. Slowly-sheared granular material experiments have revealed avalanches with power-law statistics, but currently these studies have been limited to  stationary flow \cite{Hayman11,Bretz06} (which, in addition to miss the transient behavior, may lead to additional complexity for granular materials due to the emergence of isostaticity \cite{DeGiuli15,Lerner12a}, a property however absent in the transient \cite{Kruyt10}). 
Note that for stresses far from $\Sigma_{\max}$, a large system may be required to test our views, since one must have $\langle S\rangle\gg 1$ for Eq.(\ref{01}) to hold.  Our predictions may also apply in disordered crystals, where $\Sigma_{\max}$ however is not well-defined presumably due to work hardening \cite{Miguel01,Csikor2007}. In \cite{Zapperi2014}, the authors  observe numerically scale-free avalanches with $\langle S\rangle \sim N^{0.4}$ for a range of stresses. In our views that corresponds to $\theta\approx 0.67$, a prediction that could be tested by measuring how the characteristic interval of stress with no plasticity vanishes with $N$.
Finally, a central question for the future is what governs the value of the exponent $\theta$, which affects plasticity but also macroscopic rheological properties. 

{\bf Acknowledgments}: It is a pleasure to thank E. DeGiuli,  E. Lerner, B. Metzger, M. Muller,   J. Weiss, L. Yan and  S. Zapperi for discussions related to this work. MW acknowledges support from NSF CBET Grant 1236378 and MRSEC Program of the NSF DMR-0820341 for partial funding. 
\bibliography{Wyartbibsaveapril30}

\begin{thebibliography}{41}%
\makeatletter
\providecommand \@ifxundefined [1]{%
 \@ifx{#1\undefined}
}%
\providecommand \@ifnum [1]{%
 \ifnum #1\expandafter \@firstoftwo
 \else \expandafter \@secondoftwo
 \fi
}%
\providecommand \@ifx [1]{%
 \ifx #1\expandafter \@firstoftwo
 \else \expandafter \@secondoftwo
 \fi
}%
\providecommand \natexlab [1]{#1}%
\providecommand \enquote  [1]{``#1''}%
\providecommand \bibnamefont  [1]{#1}%
\providecommand \bibfnamefont [1]{#1}%
\providecommand \citenamefont [1]{#1}%
\providecommand \href@noop [0]{\@secondoftwo}%
\providecommand \href [0]{\begingroup \@sanitize@url \@href}%
\providecommand \@href[1]{\@@startlink{#1}\@@href}%
\providecommand \@@href[1]{\endgroup#1\@@endlink}%
\providecommand \@sanitize@url [0]{\catcode `\\12\catcode `\$12\catcode
  `\&12\catcode `\#12\catcode `\^12\catcode `\_12\catcode `\%12\relax}%
\providecommand \@@startlink[1]{}%
\providecommand \@@endlink[0]{}%
\providecommand \url  [0]{\begingroup\@sanitize@url \@url }%
\providecommand \@url [1]{\endgroup\@href {#1}{\urlprefix }}%
\providecommand \urlprefix  [0]{URL }%
\providecommand \Eprint [0]{\href }%
\providecommand \doibase [0]{http://dx.doi.org/}%
\providecommand \selectlanguage [0]{\@gobble}%
\providecommand \bibinfo  [0]{\@secondoftwo}%
\providecommand \bibfield  [0]{\@secondoftwo}%
\providecommand \translation [1]{[#1]}%
\providecommand \BibitemOpen [0]{}%
\providecommand \bibitemStop [0]{}%
\providecommand \bibitemNoStop [0]{.\EOS\space}%
\providecommand \EOS [0]{\spacefactor3000\relax}%
\providecommand \BibitemShut  [1]{\csname bibitem#1\endcsname}%
\let\auto@bib@innerbib\@empty
\bibitem [{\citenamefont {Manning}\ \emph {et~al.}(2007)\citenamefont
  {Manning}, \citenamefont {Langer},\ and\ \citenamefont
  {Carlson}}]{Manning07}%
  \BibitemOpen
  \bibfield  {author} {\bibinfo {author} {\bibfnamefont {M.~L.}\ \bibnamefont
  {Manning}}, \bibinfo {author} {\bibfnamefont {J.~S.}\ \bibnamefont {Langer}},
  \ and\ \bibinfo {author} {\bibfnamefont {J.~M.}\ \bibnamefont {Carlson}},\
  }\href@noop {} {\bibfield  {journal} {\bibinfo  {journal} {{Physical review
  E}}\ }\textbf {\bibinfo {volume} {{76}}} (\bibinfo {year}
  {{2007}})}\BibitemShut {NoStop}%
\bibitem [{\citenamefont {Andreotti}\ \emph {et~al.}(2013)\citenamefont
  {Andreotti}, \citenamefont {Forterre},\ and\ \citenamefont
  {Pouliquen}}]{Andreotti13}%
  \BibitemOpen
  \bibfield  {author} {\bibinfo {author} {\bibfnamefont {B.}~\bibnamefont
  {Andreotti}}, \bibinfo {author} {\bibfnamefont {Y.}~\bibnamefont {Forterre}},
  \ and\ \bibinfo {author} {\bibfnamefont {O.}~\bibnamefont {Pouliquen}},\
  }\href@noop {} {\emph {\bibinfo {title} {Granular media: between fluid and
  solid}}}\ (\bibinfo  {publisher} {Cambridge University Press},\ \bibinfo
  {year} {2013})\BibitemShut {NoStop}%
\bibitem [{\citenamefont {Jagla}(2007)}]{Jagla07}%
  \BibitemOpen
  \bibfield  {author} {\bibinfo {author} {\bibfnamefont {E.~A.}\ \bibnamefont
  {Jagla}},\ }\href {\doibase 10.1103/PhysRevE.76.046119} {\bibfield  {journal}
  {\bibinfo  {journal} {Phys. Rev. E}\ }\textbf {\bibinfo {volume} {76}},\
  \bibinfo {pages} {046119} (\bibinfo {year} {2007})}\BibitemShut {NoStop}%
\bibitem [{\citenamefont {Divoux}\ \emph {et~al.}(2015)\citenamefont {Divoux},
  \citenamefont {Fardin}, \citenamefont {Manneville},\ and\ \citenamefont
  {Lerouge}}]{Divoux15}%
  \BibitemOpen
  \bibfield  {author} {\bibinfo {author} {\bibfnamefont {T.}~\bibnamefont
  {Divoux}}, \bibinfo {author} {\bibfnamefont {M.~A.}\ \bibnamefont {Fardin}},
  \bibinfo {author} {\bibfnamefont {S.}~\bibnamefont {Manneville}}, \ and\
  \bibinfo {author} {\bibfnamefont {S.}~\bibnamefont {Lerouge}},\ }\href@noop
  {} {\bibfield  {journal} {\bibinfo  {journal} {arXiv preprint
  arXiv:1503.04130}\ } (\bibinfo {year} {2015})}\BibitemShut {NoStop}%
\bibitem [{\citenamefont {Moorcroft}\ and\ \citenamefont
  {Fielding}(2013)}]{Moorcroft13}%
  \BibitemOpen
  \bibfield  {author} {\bibinfo {author} {\bibfnamefont {R.~L.}\ \bibnamefont
  {Moorcroft}}\ and\ \bibinfo {author} {\bibfnamefont {S.~M.}\ \bibnamefont
  {Fielding}},\ }\href@noop {} {\bibfield  {journal} {\bibinfo  {journal}
  {Physical review letters}\ }\textbf {\bibinfo {volume} {110}},\ \bibinfo
  {pages} {086001} (\bibinfo {year} {2013})}\BibitemShut {NoStop}%
\bibitem [{\citenamefont {Lockner}\ and\ \citenamefont
  {Beeler}(2002)}]{Lockner02}%
  \BibitemOpen
  \bibfield  {author} {\bibinfo {author} {\bibfnamefont {D.~A.}\ \bibnamefont
  {Lockner}}\ and\ \bibinfo {author} {\bibfnamefont {N.~M.}\ \bibnamefont
  {Beeler}},\ }\href@noop {} {\bibfield  {journal} {\bibinfo  {journal}
  {International Geophysics}\ }\textbf {\bibinfo {volume} {81}},\ \bibinfo
  {pages} {505} (\bibinfo {year} {2002})}\BibitemShut {NoStop}%
\bibitem [{\citenamefont {Amon}\ \emph {et~al.}(2012)\citenamefont {Amon},
  \citenamefont {Nguyen}, \citenamefont {Bruand}, \citenamefont {Crassous},\
  and\ \citenamefont {Cl\'ement}}]{Amon2012}%
  \BibitemOpen
  \bibfield  {author} {\bibinfo {author} {\bibfnamefont {A.}~\bibnamefont
  {Amon}}, \bibinfo {author} {\bibfnamefont {V.~B.}\ \bibnamefont {Nguyen}},
  \bibinfo {author} {\bibfnamefont {A.}~\bibnamefont {Bruand}}, \bibinfo
  {author} {\bibfnamefont {J.}~\bibnamefont {Crassous}}, \ and\ \bibinfo
  {author} {\bibfnamefont {E.}~\bibnamefont {Cl\'ement}},\ }\href {\doibase
  10.1103/PhysRevLett.108.135502} {\bibfield  {journal} {\bibinfo  {journal}
  {Phys. Rev. Lett.}\ }\textbf {\bibinfo {volume} {108}},\ \bibinfo {pages}
  {135502} (\bibinfo {year} {2012})}\BibitemShut {NoStop}%
\bibitem [{\citenamefont {Le~Bouil}\ \emph
  {et~al.}(2014{\natexlab{a}})\citenamefont {Le~Bouil}, \citenamefont {Amon},
  \citenamefont {Sangleboeuf}, \citenamefont {Orain}, \citenamefont
  {B{\'e}suelle}, \citenamefont {Viggiani}, \citenamefont {Chasle},\ and\
  \citenamefont {Crassous}}]{Lebouil2014}%
  \BibitemOpen
  \bibfield  {author} {\bibinfo {author} {\bibfnamefont {A.}~\bibnamefont
  {Le~Bouil}}, \bibinfo {author} {\bibfnamefont {A.}~\bibnamefont {Amon}},
  \bibinfo {author} {\bibfnamefont {J.-C.}\ \bibnamefont {Sangleboeuf}},
  \bibinfo {author} {\bibfnamefont {H.}~\bibnamefont {Orain}}, \bibinfo
  {author} {\bibfnamefont {P.}~\bibnamefont {B{\'e}suelle}}, \bibinfo {author}
  {\bibfnamefont {G.}~\bibnamefont {Viggiani}}, \bibinfo {author}
  {\bibfnamefont {P.}~\bibnamefont {Chasle}}, \ and\ \bibinfo {author}
  {\bibfnamefont {J.}~\bibnamefont {Crassous}},\ }\href@noop {} {\bibfield
  {journal} {\bibinfo  {journal} {Granular Matter}\ }\textbf {\bibinfo {volume}
  {16}},\ \bibinfo {pages} {1} (\bibinfo {year}
  {2014}{\natexlab{a}})}\BibitemShut {NoStop}%
\bibitem [{\citenamefont {Le~Bouil}\ \emph
  {et~al.}(2014{\natexlab{b}})\citenamefont {Le~Bouil}, \citenamefont {Amon},
  \citenamefont {McNamara},\ and\ \citenamefont {Crassous}}]{Lebouil14a}%
  \BibitemOpen
  \bibfield  {author} {\bibinfo {author} {\bibfnamefont {A.}~\bibnamefont
  {Le~Bouil}}, \bibinfo {author} {\bibfnamefont {A.}~\bibnamefont {Amon}},
  \bibinfo {author} {\bibfnamefont {S.}~\bibnamefont {McNamara}}, \ and\
  \bibinfo {author} {\bibfnamefont {J.}~\bibnamefont {Crassous}},\ }\href@noop
  {} {\bibfield  {journal} {\bibinfo  {journal} {Physical review letters}\
  }\textbf {\bibinfo {volume} {112}},\ \bibinfo {pages} {246001} (\bibinfo
  {year} {2014}{\natexlab{b}})}\BibitemShut {NoStop}%
\bibitem [{\citenamefont {Gimbert}\ \emph {et~al.}(2013)\citenamefont
  {Gimbert}, \citenamefont {Amitrano},\ and\ \citenamefont
  {Weiss}}]{Gimbert2013}%
  \BibitemOpen
  \bibfield  {author} {\bibinfo {author} {\bibfnamefont {F.}~\bibnamefont
  {Gimbert}}, \bibinfo {author} {\bibfnamefont {D.}~\bibnamefont {Amitrano}}, \
  and\ \bibinfo {author} {\bibfnamefont {J.}~\bibnamefont {Weiss}},\
  }\href@noop {} {\bibfield  {journal} {\bibinfo  {journal} {EPL (Europhysics
  Letters)}\ }\textbf {\bibinfo {volume} {104}},\ \bibinfo {pages} {46001}
  (\bibinfo {year} {2013})}\BibitemShut {NoStop}%
\bibitem [{\citenamefont {Argon}(1979)}]{Argon79}%
  \BibitemOpen
  \bibfield  {author} {\bibinfo {author} {\bibfnamefont {A.}~\bibnamefont
  {Argon}},\ }\href {\doibase 10.1016/0001-6160(79)90055-5} {\bibfield
  {journal} {\bibinfo  {journal} {Acta Metallurgica}\ }\textbf {\bibinfo
  {volume} {27}},\ \bibinfo {pages} {47 } (\bibinfo {year} {1979})}\BibitemShut
  {NoStop}%
\bibitem [{\citenamefont {Pouliquen}(2004)}]{Pouliquen04}%
  \BibitemOpen
  \bibfield  {author} {\bibinfo {author} {\bibfnamefont {O.}~\bibnamefont
  {Pouliquen}},\ }\href@noop {} {\bibfield  {journal} {\bibinfo  {journal}
  {Physical review letters}\ }\textbf {\bibinfo {volume} {93}},\ \bibinfo
  {pages} {248001} (\bibinfo {year} {2004})}\BibitemShut {NoStop}%
\bibitem [{\citenamefont {Lin}\ \emph {et~al.}(2014{\natexlab{a}})\citenamefont
  {Lin}, \citenamefont {Lerner}, \citenamefont {Rosso},\ and\ \citenamefont
  {Wyart}}]{Lin2014}%
  \BibitemOpen
  \bibfield  {author} {\bibinfo {author} {\bibfnamefont {J.}~\bibnamefont
  {Lin}}, \bibinfo {author} {\bibfnamefont {E.}~\bibnamefont {Lerner}},
  \bibinfo {author} {\bibfnamefont {A.}~\bibnamefont {Rosso}}, \ and\ \bibinfo
  {author} {\bibfnamefont {M.}~\bibnamefont {Wyart}},\ }\href@noop {}
  {\bibfield  {journal} {\bibinfo  {journal} {Proceedings of the National
  Academy of Sciences}\ }\textbf {\bibinfo {volume} {111}},\ \bibinfo {pages}
  {14382} (\bibinfo {year} {2014}{\natexlab{a}})}\BibitemShut {NoStop}%
\bibitem [{\citenamefont {{Olsson}}\ and\ \citenamefont
  {{Teitel}}(2007)}]{Olsson07}%
  \BibitemOpen
  \bibfield  {author} {\bibinfo {author} {\bibfnamefont {P.}~\bibnamefont
  {{Olsson}}}\ and\ \bibinfo {author} {\bibfnamefont {S.}~\bibnamefont
  {{Teitel}}},\ }\href@noop {} {\bibfield  {journal} {\bibinfo  {journal}
  {Phys.\ Rev.\ Lett.}\ }\textbf {\bibinfo {volume} {99}},\ \bibinfo {pages}
  {178001} (\bibinfo {year} {2007})}\BibitemShut {NoStop}%
\bibitem [{\citenamefont {Martens}\ \emph {et~al.}(2011)\citenamefont
  {Martens}, \citenamefont {Bocquet},\ and\ \citenamefont
  {Barrat}}]{Martens11}%
  \BibitemOpen
  \bibfield  {author} {\bibinfo {author} {\bibfnamefont {K.}~\bibnamefont
  {Martens}}, \bibinfo {author} {\bibfnamefont {L.}~\bibnamefont {Bocquet}}, \
  and\ \bibinfo {author} {\bibfnamefont {J.-L.}\ \bibnamefont {Barrat}},\
  }\href {\doibase 10.1103/PhysRevLett.106.156001} {\bibfield  {journal}
  {\bibinfo  {journal} {Phys. Rev. Lett.}\ }\textbf {\bibinfo {volume} {106}},\
  \bibinfo {pages} {156001} (\bibinfo {year} {2011})}\BibitemShut {NoStop}%
\bibitem [{\citenamefont {D{\"u}ring}\ \emph {et~al.}(2014)\citenamefont
  {D{\"u}ring}, \citenamefont {Lerner},\ and\ \citenamefont
  {Wyart}}]{During14}%
  \BibitemOpen
  \bibfield  {author} {\bibinfo {author} {\bibfnamefont {G.}~\bibnamefont
  {D{\"u}ring}}, \bibinfo {author} {\bibfnamefont {E.}~\bibnamefont {Lerner}},
  \ and\ \bibinfo {author} {\bibfnamefont {M.}~\bibnamefont {Wyart}},\
  }\href@noop {} {\bibfield  {journal} {\bibinfo  {journal} {Physical Review
  E}\ }\textbf {\bibinfo {volume} {89}},\ \bibinfo {pages} {022305} (\bibinfo
  {year} {2014})}\BibitemShut {NoStop}%
\bibitem [{\citenamefont {Salerno}\ \emph {et~al.}(2012)\citenamefont
  {Salerno}, \citenamefont {Maloney},\ and\ \citenamefont
  {Robbins}}]{Salerno12}%
  \BibitemOpen
  \bibfield  {author} {\bibinfo {author} {\bibfnamefont {K.~M.}\ \bibnamefont
  {Salerno}}, \bibinfo {author} {\bibfnamefont {C.~E.}\ \bibnamefont
  {Maloney}}, \ and\ \bibinfo {author} {\bibfnamefont {M.~O.}\ \bibnamefont
  {Robbins}},\ }\href {\doibase 10.1103/PhysRevLett.109.105703} {\bibfield
  {journal} {\bibinfo  {journal} {Phys. Rev. Lett.}\ }\textbf {\bibinfo
  {volume} {109}},\ \bibinfo {pages} {105703} (\bibinfo {year}
  {2012})}\BibitemShut {NoStop}%
\bibitem [{\citenamefont {Lema\^itre}\ and\ \citenamefont
  {Caroli}(2009)}]{Lemaitre09}%
  \BibitemOpen
  \bibfield  {author} {\bibinfo {author} {\bibfnamefont {A.}~\bibnamefont
  {Lema\^itre}}\ and\ \bibinfo {author} {\bibfnamefont {C.}~\bibnamefont
  {Caroli}},\ }\href {\doibase 10.1103/PhysRevLett.103.065501} {\bibfield
  {journal} {\bibinfo  {journal} {Phys. Rev. Lett.}\ }\textbf {\bibinfo
  {volume} {103}},\ \bibinfo {pages} {065501} (\bibinfo {year}
  {2009})}\BibitemShut {NoStop}%
\bibitem [{\citenamefont {Fisher}(1998)}]{Fisher98}%
  \BibitemOpen
  \bibfield  {author} {\bibinfo {author} {\bibfnamefont {D.~S.}\ \bibnamefont
  {Fisher}},\ }\href {\doibase 10.1016/S0370-1573(98)00008-8} {\bibfield
  {journal} {\bibinfo  {journal} {Physics Reports}\ }\textbf {\bibinfo {volume}
  {301}},\ \bibinfo {pages} {113 } (\bibinfo {year} {1998})}\BibitemShut
  {NoStop}%
\bibitem [{\citenamefont {M{\"u}ller}\ and\ \citenamefont
  {Wyart}(2015)}]{Muller14}%
  \BibitemOpen
  \bibfield  {author} {\bibinfo {author} {\bibfnamefont {M.}~\bibnamefont
  {M{\"u}ller}}\ and\ \bibinfo {author} {\bibfnamefont {M.}~\bibnamefont
  {Wyart}},\ }\href@noop {} {\bibfield  {journal} {\bibinfo  {journal} {Annual
  Review of Condensed Matter Physics}\ }\textbf {\bibinfo {volume} {6}}
  (\bibinfo {year} {2015})}\BibitemShut {NoStop}%
\bibitem [{\citenamefont {Lin}\ \emph {et~al.}(2014{\natexlab{b}})\citenamefont
  {Lin}, \citenamefont {Saade}, \citenamefont {Lerner}, \citenamefont {Rosso},\
  and\ \citenamefont {Wyart}}]{Lin14a}%
  \BibitemOpen
  \bibfield  {author} {\bibinfo {author} {\bibfnamefont {J.}~\bibnamefont
  {Lin}}, \bibinfo {author} {\bibfnamefont {A.}~\bibnamefont {Saade}}, \bibinfo
  {author} {\bibfnamefont {E.}~\bibnamefont {Lerner}}, \bibinfo {author}
  {\bibfnamefont {A.}~\bibnamefont {Rosso}}, \ and\ \bibinfo {author}
  {\bibfnamefont {M.}~\bibnamefont {Wyart}},\ }\href@noop {} {\bibfield
  {journal} {\bibinfo  {journal} {EPL (Europhysics Letters)}\ }\textbf
  {\bibinfo {volume} {105}},\ \bibinfo {pages} {26003} (\bibinfo {year}
  {2014}{\natexlab{b}})}\BibitemShut {NoStop}%
\bibitem [{\citenamefont {Picard}\ \emph {et~al.}(2005)\citenamefont {Picard},
  \citenamefont {Ajdari}, \citenamefont {Lequeux},\ and\ \citenamefont
  {Bocquet}}]{Picard2005}%
  \BibitemOpen
  \bibfield  {author} {\bibinfo {author} {\bibfnamefont {G.}~\bibnamefont
  {Picard}}, \bibinfo {author} {\bibfnamefont {A.}~\bibnamefont {Ajdari}},
  \bibinfo {author} {\bibfnamefont {F.}~\bibnamefont {Lequeux}}, \ and\
  \bibinfo {author} {\bibfnamefont {L.}~\bibnamefont {Bocquet}},\ }\href@noop
  {} {\bibfield  {journal} {\bibinfo  {journal} {Physical Review E}\ }\textbf
  {\bibinfo {volume} {71}},\ \bibinfo {pages} {010501} (\bibinfo {year}
  {2005})}\BibitemShut {NoStop}%
\bibitem [{\citenamefont {Baret}\ \emph {et~al.}(2002)\citenamefont {Baret},
  \citenamefont {Vandembroucq},\ and\ \citenamefont {Roux}}]{Baret02}%
  \BibitemOpen
  \bibfield  {author} {\bibinfo {author} {\bibfnamefont {J.-C.}\ \bibnamefont
  {Baret}}, \bibinfo {author} {\bibfnamefont {D.}~\bibnamefont {Vandembroucq}},
  \ and\ \bibinfo {author} {\bibfnamefont {S.}~\bibnamefont {Roux}},\ }\href
  {\doibase 10.1103/PhysRevLett.89.195506} {\bibfield  {journal} {\bibinfo
  {journal} {Phys. Rev. Lett.}\ }\textbf {\bibinfo {volume} {89}},\ \bibinfo
  {pages} {195506} (\bibinfo {year} {2002})}\BibitemShut {NoStop}%
\bibitem [{\citenamefont {H\'ebraud}\ and\ \citenamefont
  {Lequeux}(1998)}]{Hebraud98}%
  \BibitemOpen
  \bibfield  {author} {\bibinfo {author} {\bibfnamefont {P.}~\bibnamefont
  {H\'ebraud}}\ and\ \bibinfo {author} {\bibfnamefont {F.}~\bibnamefont
  {Lequeux}},\ }\href {\doibase 10.1103/PhysRevLett.81.2934} {\bibfield
  {journal} {\bibinfo  {journal} {Phys. Rev. Lett.}\ }\textbf {\bibinfo
  {volume} {81}},\ \bibinfo {pages} {2934} (\bibinfo {year}
  {1998})}\BibitemShut {NoStop}%
\bibitem [{\citenamefont {Picard}\ \emph {et~al.}(2004)\citenamefont {Picard},
  \citenamefont {Ajdari}, \citenamefont {Lequeux},\ and\ \citenamefont
  {Bocquet}}]{Picard04}%
  \BibitemOpen
  \bibfield  {author} {\bibinfo {author} {\bibfnamefont {G.}~\bibnamefont
  {Picard}}, \bibinfo {author} {\bibfnamefont {A.}~\bibnamefont {Ajdari}},
  \bibinfo {author} {\bibfnamefont {F.}~\bibnamefont {Lequeux}}, \ and\
  \bibinfo {author} {\bibfnamefont {L.}~\bibnamefont {Bocquet}},\ }\href
  {\doibase 10.1140/epje/i2004-10054-8} {\bibfield  {journal} {\bibinfo
  {journal} {The European Physical Journal E}\ }\textbf {\bibinfo {volume}
  {15}},\ \bibinfo {pages} {371} (\bibinfo {year} {2004})}\BibitemShut
  {NoStop}%
\bibitem [{\citenamefont {Vandembroucq}\ and\ \citenamefont
  {Roux}(2011)}]{Vandembroucq2011}%
  \BibitemOpen
  \bibfield  {author} {\bibinfo {author} {\bibfnamefont {D.}~\bibnamefont
  {Vandembroucq}}\ and\ \bibinfo {author} {\bibfnamefont {S.}~\bibnamefont
  {Roux}},\ }\href@noop {} {\bibfield  {journal} {\bibinfo  {journal} {Physical
  Review B}\ }\textbf {\bibinfo {volume} {84}},\ \bibinfo {pages} {134210}
  (\bibinfo {year} {2011})}\BibitemShut {NoStop}%
\bibitem [{\citenamefont {Shi}\ \emph {et~al.}(2007)\citenamefont {Shi},
  \citenamefont {Katz}, \citenamefont {Li},\ and\ \citenamefont
  {Falk}}]{Shi2007}%
  \BibitemOpen
  \bibfield  {author} {\bibinfo {author} {\bibfnamefont {Y.}~\bibnamefont
  {Shi}}, \bibinfo {author} {\bibfnamefont {M.~B.}\ \bibnamefont {Katz}},
  \bibinfo {author} {\bibfnamefont {H.}~\bibnamefont {Li}}, \ and\ \bibinfo
  {author} {\bibfnamefont {M.~L.}\ \bibnamefont {Falk}},\ }\href@noop {}
  {\bibfield  {journal} {\bibinfo  {journal} {Physical review letters}\
  }\textbf {\bibinfo {volume} {98}},\ \bibinfo {pages} {185505} (\bibinfo
  {year} {2007})}\BibitemShut {NoStop}%
\bibitem [{\citenamefont {Moorcroft}\ \emph {et~al.}(2011)\citenamefont
  {Moorcroft}, \citenamefont {Cates},\ and\ \citenamefont
  {Fielding}}]{Moorcroft2011}%
  \BibitemOpen
  \bibfield  {author} {\bibinfo {author} {\bibfnamefont {R.~L.}\ \bibnamefont
  {Moorcroft}}, \bibinfo {author} {\bibfnamefont {M.~E.}\ \bibnamefont
  {Cates}}, \ and\ \bibinfo {author} {\bibfnamefont {S.~M.}\ \bibnamefont
  {Fielding}},\ }\href@noop {} {\bibfield  {journal} {\bibinfo  {journal}
  {Physical review letters}\ }\textbf {\bibinfo {volume} {106}},\ \bibinfo
  {pages} {055502} (\bibinfo {year} {2011})}\BibitemShut {NoStop}%
\bibitem [{\citenamefont {Fielding}(2014)}]{Fielding2014}%
  \BibitemOpen
  \bibfield  {author} {\bibinfo {author} {\bibfnamefont {S.~M.}\ \bibnamefont
  {Fielding}},\ }\href@noop {} {\bibfield  {journal} {\bibinfo  {journal}
  {Reports on Progress in Physics}\ }\textbf {\bibinfo {volume} {77}},\
  \bibinfo {pages} {102601} (\bibinfo {year} {2014})}\BibitemShut {NoStop}%
\bibitem [{\citenamefont {Lema{\^\i}tre}\ and\ \citenamefont
  {Caroli}(2007)}]{Lemaitre07}%
  \BibitemOpen
  \bibfield  {author} {\bibinfo {author} {\bibfnamefont {A.}~\bibnamefont
  {Lema{\^\i}tre}}\ and\ \bibinfo {author} {\bibfnamefont {C.}~\bibnamefont
  {Caroli}},\ }\href@noop {} {\bibfield  {journal} {\bibinfo  {journal} {arXiv
  preprint arXiv:0705.3122}\ } (\bibinfo {year} {2007})}\BibitemShut {NoStop}%
\bibitem [{\citenamefont {Karmakar}\ \emph {et~al.}(2010)\citenamefont
  {Karmakar}, \citenamefont {Lerner},\ and\ \citenamefont
  {Procaccia}}]{Karmakar10a}%
  \BibitemOpen
  \bibfield  {author} {\bibinfo {author} {\bibfnamefont {S.}~\bibnamefont
  {Karmakar}}, \bibinfo {author} {\bibfnamefont {E.}~\bibnamefont {Lerner}}, \
  and\ \bibinfo {author} {\bibfnamefont {I.}~\bibnamefont {Procaccia}},\ }\href
  {\doibase 10.1103/PhysRevE.82.055103} {\bibfield  {journal} {\bibinfo
  {journal} {Phys. Rev. E}\ }\textbf {\bibinfo {volume} {82}},\ \bibinfo
  {pages} {055103} (\bibinfo {year} {2010})}\BibitemShut {NoStop}%
\bibitem [{\citenamefont {Salerno}\ and\ \citenamefont
  {Robbins}(2013)}]{Salerno13}%
  \BibitemOpen
  \bibfield  {author} {\bibinfo {author} {\bibfnamefont {K.~M.}\ \bibnamefont
  {Salerno}}\ and\ \bibinfo {author} {\bibfnamefont {M.~O.}\ \bibnamefont
  {Robbins}},\ }\href@noop {} {\bibfield  {journal} {\bibinfo  {journal}
  {Physical Review E}\ }\textbf {\bibinfo {volume} {88}},\ \bibinfo {pages}
  {062206} (\bibinfo {year} {2013})}\BibitemShut {NoStop}%
\bibitem [{\citenamefont {Girard}\ \emph {et~al.}(2010)\citenamefont {Girard},
  \citenamefont {Amitrano},\ and\ \citenamefont {Weiss}}]{Girard2010}%
  \BibitemOpen
  \bibfield  {author} {\bibinfo {author} {\bibfnamefont {L.}~\bibnamefont
  {Girard}}, \bibinfo {author} {\bibfnamefont {D.}~\bibnamefont {Amitrano}}, \
  and\ \bibinfo {author} {\bibfnamefont {J.}~\bibnamefont {Weiss}},\
  }\href@noop {} {\bibfield  {journal} {\bibinfo  {journal} {Journal of
  Statistical Mechanics: Theory and Experiment}\ }\textbf {\bibinfo {volume}
  {2010}},\ \bibinfo {pages} {P01013} (\bibinfo {year} {2010})}\BibitemShut
  {NoStop}%
\bibitem [{\citenamefont {Hayman}\ \emph {et~al.}(2011)\citenamefont {Hayman},
  \citenamefont {Duclou{\'e}}, \citenamefont {Foco},\ and\ \citenamefont
  {Daniels}}]{Hayman11}%
  \BibitemOpen
  \bibfield  {author} {\bibinfo {author} {\bibfnamefont {N.~W.}\ \bibnamefont
  {Hayman}}, \bibinfo {author} {\bibfnamefont {L.}~\bibnamefont {Duclou{\'e}}},
  \bibinfo {author} {\bibfnamefont {K.~L.}\ \bibnamefont {Foco}}, \ and\
  \bibinfo {author} {\bibfnamefont {K.~E.}\ \bibnamefont {Daniels}},\
  }\href@noop {} {\bibfield  {journal} {\bibinfo  {journal} {Pure and applied
  geophysics}\ }\textbf {\bibinfo {volume} {168}},\ \bibinfo {pages} {2239}
  (\bibinfo {year} {2011})}\BibitemShut {NoStop}%
\bibitem [{\citenamefont {Bretz}\ \emph {et~al.}(2006)\citenamefont {Bretz},
  \citenamefont {Zaretzki}, \citenamefont {Field}, \citenamefont {Mitarai},\
  and\ \citenamefont {Nori}}]{Bretz06}%
  \BibitemOpen
  \bibfield  {author} {\bibinfo {author} {\bibfnamefont {M.}~\bibnamefont
  {Bretz}}, \bibinfo {author} {\bibfnamefont {R.}~\bibnamefont {Zaretzki}},
  \bibinfo {author} {\bibfnamefont {S.~B.}\ \bibnamefont {Field}}, \bibinfo
  {author} {\bibfnamefont {N.}~\bibnamefont {Mitarai}}, \ and\ \bibinfo
  {author} {\bibfnamefont {F.}~\bibnamefont {Nori}},\ }\href@noop {} {\bibfield
   {journal} {\bibinfo  {journal} {EPL (Europhysics Letters)}\ }\textbf
  {\bibinfo {volume} {74}},\ \bibinfo {pages} {1116} (\bibinfo {year}
  {2006})}\BibitemShut {NoStop}%
\bibitem [{\citenamefont {DeGiuli}\ \emph {et~al.}(2015)\citenamefont
  {DeGiuli}, \citenamefont {D\"uring}, \citenamefont {Lerner},\ and\
  \citenamefont {Wyart}}]{DeGiuli15}%
  \BibitemOpen
  \bibfield  {author} {\bibinfo {author} {\bibfnamefont {E.}~\bibnamefont
  {DeGiuli}}, \bibinfo {author} {\bibfnamefont {G.}~\bibnamefont {D\"uring}},
  \bibinfo {author} {\bibfnamefont {E.}~\bibnamefont {Lerner}}, \ and\ \bibinfo
  {author} {\bibfnamefont {M.}~\bibnamefont {Wyart}},\ }\href {\doibase
  10.1103/PhysRevE.91.062206} {\bibfield  {journal} {\bibinfo  {journal} {Phys.
  Rev. E}\ }\textbf {\bibinfo {volume} {91}},\ \bibinfo {pages} {062206}
  (\bibinfo {year} {2015})}\BibitemShut {NoStop}%
\bibitem [{\citenamefont {Lerner}\ \emph {et~al.}(2012)\citenamefont {Lerner},
  \citenamefont {D\"uring},\ and\ \citenamefont {Wyart}}]{Lerner12a}%
  \BibitemOpen
  \bibfield  {author} {\bibinfo {author} {\bibfnamefont {E.}~\bibnamefont
  {Lerner}}, \bibinfo {author} {\bibfnamefont {G.}~\bibnamefont {D\"uring}}, \
  and\ \bibinfo {author} {\bibfnamefont {M.}~\bibnamefont {Wyart}},\
  }\href@noop {} {\bibfield  {journal} {\bibinfo  {journal} {Proceedings of the
  National Academy of Sciences}\ }\textbf {\bibinfo {volume} {109}},\ \bibinfo
  {pages} {4798} (\bibinfo {year} {2012})}\BibitemShut {NoStop}%
\bibitem [{\citenamefont {Kruyt}(2010)}]{Kruyt10}%
  \BibitemOpen
  \bibfield  {author} {\bibinfo {author} {\bibfnamefont {N.~P.}\ \bibnamefont
  {Kruyt}},\ }\href@noop {} {\bibfield  {journal} {\bibinfo  {journal} {Comptes
  Rendus M{\'e}canique}\ }\textbf {\bibinfo {volume} {338}},\ \bibinfo {pages}
  {596} (\bibinfo {year} {2010})}\BibitemShut {NoStop}%
\bibitem [{\citenamefont {Miguel}\ \emph {et~al.}(2001)\citenamefont {Miguel},
  \citenamefont {Vespignani}, \citenamefont {Zapperi}, \citenamefont {Weiss},\
  and\ \citenamefont {Grasso}}]{Miguel01}%
  \BibitemOpen
  \bibfield  {author} {\bibinfo {author} {\bibfnamefont {M.-C.}\ \bibnamefont
  {Miguel}}, \bibinfo {author} {\bibfnamefont {A.}~\bibnamefont {Vespignani}},
  \bibinfo {author} {\bibfnamefont {S.}~\bibnamefont {Zapperi}}, \bibinfo
  {author} {\bibfnamefont {J.}~\bibnamefont {Weiss}}, \ and\ \bibinfo {author}
  {\bibfnamefont {J.-R.}\ \bibnamefont {Grasso}},\ }\href {\doibase
  10.1038/35070524} {\bibfield  {journal} {\bibinfo  {journal} {Nature}\
  }\textbf {\bibinfo {volume} {410}},\ \bibinfo {pages} {667} (\bibinfo {year}
  {2001})}\BibitemShut {NoStop}%
\bibitem [{\citenamefont {Csikor}\ \emph {et~al.}(2007)\citenamefont {Csikor},
  \citenamefont {Motz}, \citenamefont {Weygand}, \citenamefont {Zaiser},\ and\
  \citenamefont {Zapperi}}]{Csikor2007}%
  \BibitemOpen
  \bibfield  {author} {\bibinfo {author} {\bibfnamefont {F.~F.}\ \bibnamefont
  {Csikor}}, \bibinfo {author} {\bibfnamefont {C.}~\bibnamefont {Motz}},
  \bibinfo {author} {\bibfnamefont {D.}~\bibnamefont {Weygand}}, \bibinfo
  {author} {\bibfnamefont {M.}~\bibnamefont {Zaiser}}, \ and\ \bibinfo {author}
  {\bibfnamefont {S.}~\bibnamefont {Zapperi}},\ }\href {\doibase
  10.1126/science.1143719} {\bibfield  {journal} {\bibinfo  {journal}
  {Science}\ }\textbf {\bibinfo {volume} {318}},\ \bibinfo {pages} {251}
  (\bibinfo {year} {2007})}\BibitemShut {NoStop}%
\bibitem [{\citenamefont {Isp\'anovity}\ \emph {et~al.}(2014)\citenamefont
  {Isp\'anovity}, \citenamefont {Laurson}, \citenamefont {Zaiser},
  \citenamefont {Groma}, \citenamefont {Zapperi},\ and\ \citenamefont
  {Alava}}]{Zapperi2014}%
  \BibitemOpen
  \bibfield  {author} {\bibinfo {author} {\bibfnamefont {P.~D.}\ \bibnamefont
  {Isp\'anovity}}, \bibinfo {author} {\bibfnamefont {L.}~\bibnamefont
  {Laurson}}, \bibinfo {author} {\bibfnamefont {M.}~\bibnamefont {Zaiser}},
  \bibinfo {author} {\bibfnamefont {I.}~\bibnamefont {Groma}}, \bibinfo
  {author} {\bibfnamefont {S.}~\bibnamefont {Zapperi}}, \ and\ \bibinfo
  {author} {\bibfnamefont {M.~J.}\ \bibnamefont {Alava}},\ }\href {\doibase
  10.1103/PhysRevLett.112.235501} {\bibfield  {journal} {\bibinfo  {journal}
  {Phys. Rev. Lett.}\ }\textbf {\bibinfo {volume} {112}},\ \bibinfo {pages}
  {235501} (\bibinfo {year} {2014})}\BibitemShut {NoStop}%
\end{thebibliography}%

\newpage
\section{Appendix}
\renewcommand{\theequation}{A.\arabic{equation}}
\setcounter{equation}{0}

\renewcommand{\thefigure}{A.\arabic{figure}}
\setcounter{figure}{0}

\subsubsection{(a) Dependence on initial conditions}

\begin{figure}[htb!]
   \includegraphics[width=.5\textwidth]{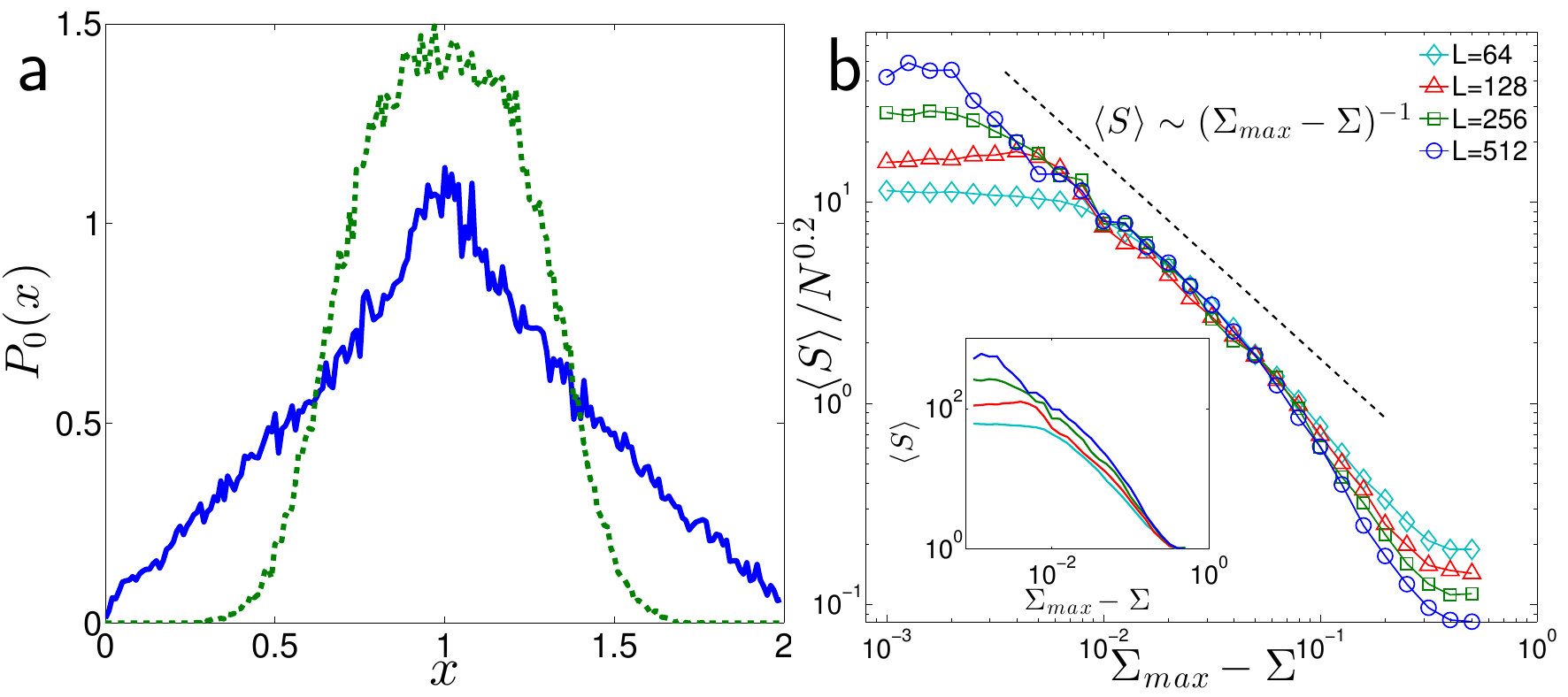} 
    \caption{(a) The initial distributions $P_0(x)$ before shear starts. Blue solid  line corresponds to $\Delta x=3.0$ (broad, no overshoot) whereas green dashed line corresponds to $\Delta x=0.7$ (narrow, overshoot). (b) Rescaled mean avalanche size $\langle S\rangle/N^{0.2}$ for the overshoot case($\Delta x=0.9$), where $\langle S\rangle$ is again observed to be size -dependent, and $\gamma\approx 1$. The dased line has a slope $-1$.}\label{figureap1}
\end{figure}

Different transient behaviors can be generated by considering different initial conditions for the distribution of  local stability $P_0(x)$, which we build as follows. We set initially $x_i=1-\sum_{j} {\cal G}(\vec r_{ij}) \Delta x_j$, where  $\Delta x_j$ is a random variable uniformly distributed in $[0, \Delta x]$. Such initial conditions ensure that each line and column present the same mean $x$, which reduces finite size effects \cite{Lin2014}. The dynamics is then run at fixed stress, until all sites are stabilized. We choose $\Delta x=3.0$ (broad), and $\Delta x=0.7$(narrow), as for Fig.1 in the main text.  The two corresponding distributions  are shown in Fig.\ref{figureap1}(a).

Our predictions, including the result of Eq.(1), appear to hold true independently of the system preparation, as shown in Fig.\ref{figureap1}(b) testing Eq.(1) in the overshoot case. Although we did not perform a detailed measurement of the exponent $\theta(\Sigma)$ in that case, we find that $\langle S\rangle$ is indeed system size dependent, and again find $\gamma\approx1$.

\subsubsection{ (b) Finite size effects on $P(x)$ and $\theta(\Sigma)$ }
\begin{figure}[H]
   \includegraphics[width=.5\textwidth]{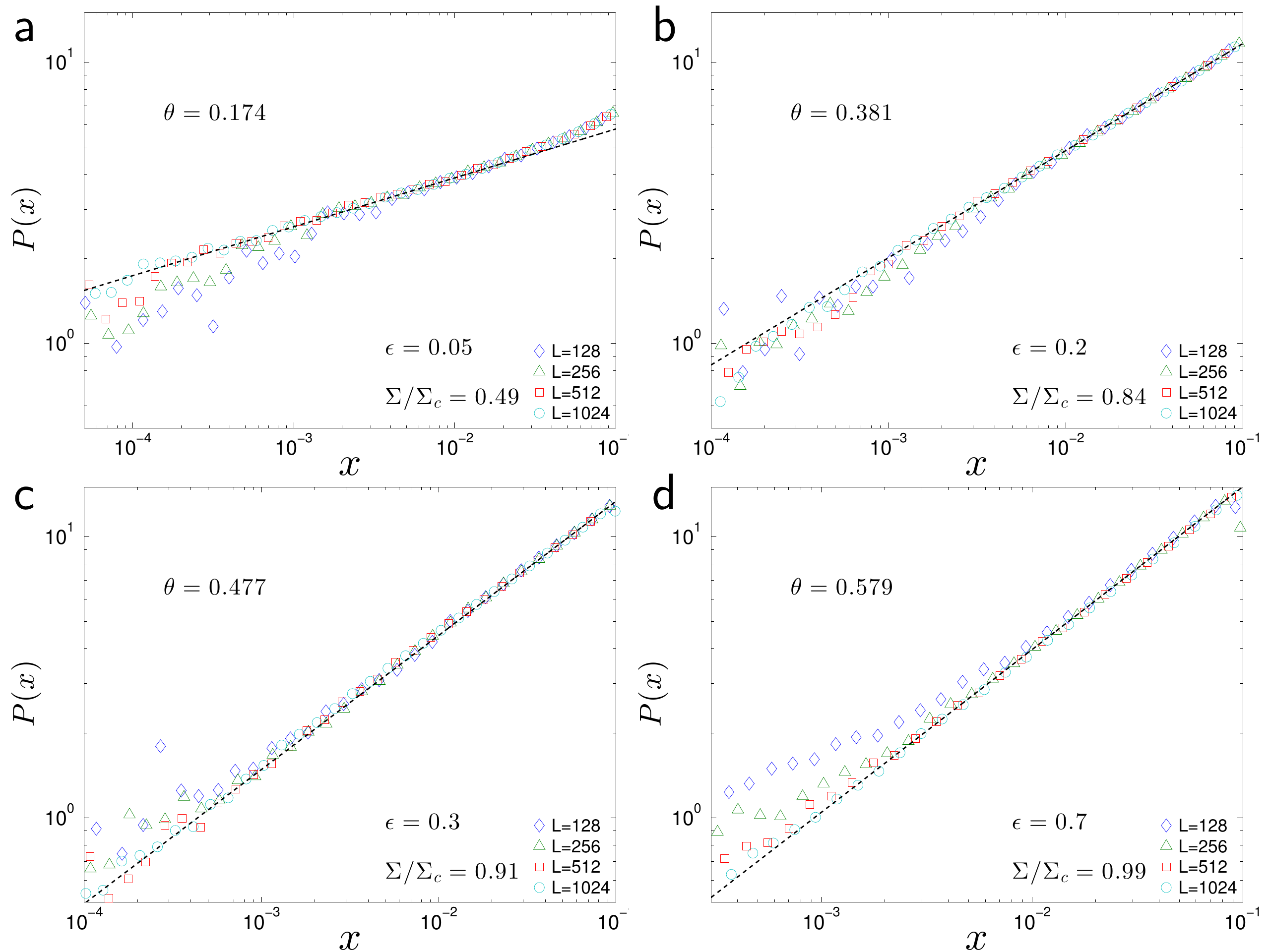} 
    \caption{$P(x)$ as a function of $\Sigma$ and system size in the stress-controlled protocol.}\label{figureap3}
\end{figure}

Fig.\ref{figureap3} shows $P(x)$ for different system sizes and four different stress values, supporting that the value we report for $\theta(\Sigma)$ in the main text, Fig.2(b), indeed approximates the true value in the thermodynamic limit. \\

%
\end{document}